\documentclass[conference]{IEEEtran}
\makeatletter
 \def\@textbottom{\vskip \z@ \@plus 0.1pt}
 \let\@texttop\relax
\makeatother
 
\usepackage{comment}
\usepackage{xspace}

 \usepackage{flushend}

\usepackage{cite}

\ifCLASSINFOpdf
  \usepackage[pdftex]{graphicx}
  \graphicspath{{../pdf/}{../jpeg/}}
  \DeclareGraphicsExtensions{.pdf,.jpeg,.png}
\else
  \usepackage[dvips]{graphicx}
  \graphicspath{{../}}
  \DeclareGraphicsExtensions{.eps}
\fi

\usepackage[cmex10]{amsmath}
 
 \usepackage{relsize} 

\usepackage[noend]{algpseudocode}
 \usepackage{array}
\ifCLASSOPTIONcompsoc
 \usepackage[caption=false,font=normalsize,labelfont=sf,textfont=sf]{subfig}
\else
 \usepackage[caption=false,font=footnotesize]{subfig}
\fi

\begin{document}
 
\title{A Highly Tunable Virtual Topology Controller}	
 
\author{\IEEEauthorblockN{Y.\,Sinan Hanay\IEEEauthorrefmark{1}, Shin'ichi Arakawa\IEEEauthorrefmark{2}\IEEEauthorrefmark{1} and Masayuki Murata\IEEEauthorrefmark{2}\IEEEauthorrefmark{1}
}
\IEEEauthorblockA{\IEEEauthorrefmark{1}
%Center for Information and Neural Networks (CiNet)\\
National Institute of Information and Communications Technology, Japan \\
Email: hanay@nict.go.jp\\ }
\IEEEauthorblockA{\IEEEauthorrefmark{2}
Graduate School of Information Science and Technology, Osaka University, Japan\\
Email: \{arakawa, murata\}@ist.osaka-u.ac.jp}}

\maketitle

\begin{abstract}
%\boldmath 

Much research in the last two decades has focused on Virtual Topology Reconfiguration (VTR) problem. However, most of the proposed methods either has low controllability, or the analysis of a control parameter is limited to empirical analysis. In this paper, we present a highly tunable Virtual Topology (VT) controller. First, we analyze the controllability of two previously proposed VTR algorithms: a heuristic method and a neural networks based method. Then we present insights on how to transform these VTR methods to their tunable versions. To benefit from the controllability, an optimality analysis  of the control parameter is needed. In the second part of the paper, through a probabilistic analysis we find an optimal parameter for the neural network based method. We validated our analysis through simulations. We propose this highly tunable method as a new VTR algorithm.
 
\end{abstract}
 

\IEEEpeerreviewmaketitle
\section{Introduction}
% no \IEEEPARstar
Optical fiber has been the main choice of communication medium for long-haul networks because of its low transmission loss.
In addition,  a fiber cable  can carry many channels simultaneously using wavelength-division multiplexing (WDM).  which makes it possible to establish
many different {\em virtual topologies} on top of the physical topology. 

Virtual topologies consist of {\em lightpaths}, which can be as short as a segment of a fiber between two hops, or 
as long as the span of sevaral fibers.
A virtual topology where each node pair is connected to each other (i.e. a complete graph) is ideal. 
However, setting up such a high degree graph may be unattainable as the number of transceivers per node is inadequate.
Instead, virtual topologies are constructed to target a performance goal such as minimizing the maximal load on any link, minimizing average hop or minimizing 
the latency between the pairs. The virtual topology reconfiguration (VTR) problem is to find a suitable topology satisfying the performance metric for the given traffic 
and resources (i.e. transceivers, number of wavelengths).

In the last decade, a lot of effort has been devoted to VTR problem in fixed WDM networks, where 
a fixed bandwidth is allocated between nodes \cite{gencata:virtual-topology,Ramaswami:DLTW96,Leonardi:ALT08}.
Elastic optical networks (EON) is a recently emerging paradigm. As opposed to fixed WDM networks,  EON proposes fine-grain bandwidth allocation that depends on traffic demand \cite{Gerstel:EON12}. 
Such a flexible physical layer requires the logical layer to be tunable as well since traffic patterns will change more frequently in near future \cite{Gerstel2013653}.
%The previously proposed VTR methods face a few challenges in the context of EONs. 
 Another major challenge in VTR problem is that heuristic methods fail to provide a ``control'' on the quality of finding a solution \cite{Leonardi:ALT08}.
 
In this work, we transformed two previously proposed VTR algorithms,
 Heuristic Logic Topology Design Algorithm (HLDA)\,\cite{Ramaswami:DLTW96} and 
 Attractor Selection Based (ASB) method\,\cite{Koizumi:10},
to their tunable versions. 
In the first part of this paper, we present our simulations on tunability of these methods.
We first start by showing that without a control parameter, these methods waste resources by  excessively establishing
lightpaths.
% A tunable method, 
% If the same performance can be achieved without depleting all resources, 
% a tunable method has to overcome this problem.
   \begin{figure*}[tb] 
\centering
  \includegraphics[width=0.9\textwidth ]{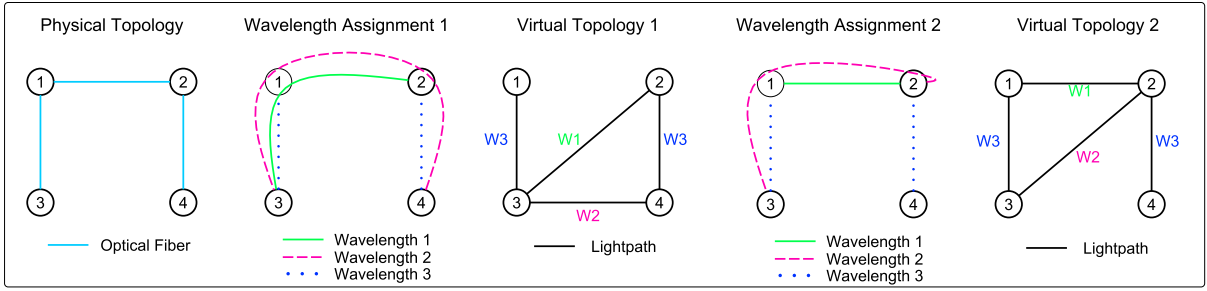}
\caption{ A virtual network topology example. In the physical topology, each physical link is a fiber, and 
    each fiber can carry three wavelengths. Both virtual topologies have 4 lightpaths. 
    Two possible virtual topologies are shown. VTR problem is to find the topology that performs better. 
 }
\label{vnt}
\end{figure*} 

In the second part of the paper, we present a probabilistic analysis of ASB to find an optimal parameter 
for adaptability to address the highly dynamic traffic. 
In order to see if our assumptions regarding the optimal parameter holds, we did simulations and
the simulations  validated our approach.

The remainder of this paper is organized as follows.% Section \ref{related} explains previous work on VTR methods.
Section \ref{problem} presents the problem setting, Section \ref{background} presents preliminaries. 
Section \ref{methodology} presents tunability and optimality analysis of the methods. 
In Section \ref{simulations} simulation results are presented and finally Section \ref{conclusion} concludes the paper.

\section{Problem Definition} \label{problem}
This paper focuses on VTR problem.
Figure \ref{vnt} illustrates the problem setting. 
A physical network consisting of four routers is given and the routers are connected through optical fiber links.
In the illustration, it is assumed that each optical fiber can carry three wavelengths. The virtual topologies have four light paths.

Wavelength assignment, lightpath and traffic routing are other aspects of virtual topology design problem \cite{zheng:OWN04}.
In this work, we only focus on VTR aspect. In other words, we are interested to find out virtual topologies that meet a performance requirement.
We assume that the routers are equipped with wavelength converters, and we use Dijsktra's shortest path algorithm for lightpath and traffic routing.

\subsection{NP-complete Problems}
VTR problem is known to be NP-complete \cite{Chlamtac:LCA92}.
Like many other NP-complete problems, exact solutions to VTR problem can be obtained using mixed integer linear programming (MILP).
%, butMILP methods become intractable for more than ten nodes.
   \begin{figure}[bht] 
\centering
  \includegraphics [width=0.8\columnwidth ]{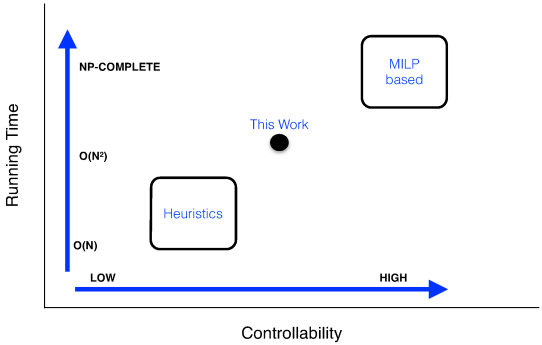}
\caption{ MILP based methods are highly tunable, however they are intractable more than about 10 nodes.
Heuristics methods are efficient, yet they cannot provide any controllability on the solution.
 }
\label{npc}
\end{figure} 
%Many efficient heuristics have been proposed earlier. 
Figure \ref{npc} summarizes the approaches to the VTR problem.
In this paper we use  terms controllability and tunability to refer that an algorithm 
 can run based on a specification. It is possible to fully specify the constraints using a MILP formulation, however
 MILP methods are intractable for more than ten nodes, and earlier work considered only topologies having less than 14 nodes \cite{gencata:virtual-topology,Ramaswami:DLTW96}.
 However, real world topologies consist of more than a few dozen nodes, for example AT\&T consists of 154 nodes and DFN network topology consists of 30 nodes \cite{Heckman:RNT03}. 
 Even a very recent work that partially uses MILP considers topologies of 6, 11 and 23 nodes\cite{Aparicio-Pardo:RVT12}.
 
 Most of the heuristic methods assign lightpaths to remaining resources
 after the algorithm finishes.
In order to evaluate different VTR algorithms, the algorithms must be evaluated also based on their overhead.
This work evaluate methods not only by their performance, but also the overhead they introduce. %We focus on two previously proposed methods:
 
 \subsection{Motivation}
Figure \ref{fixedVTR} presents the motivating example for this work. 
Three different VTR algorithms were compared by running each method 30 times for each traffic load.
 HLDA clearly performs better than  ASB and  MADN
 for traffic loads higher than $0.2$ as shown in Figure \ref{fixedPer}.
 On the other hand, Figure \ref{fixedPer} reveals that average number of lightpath change 
 per round with HLDA is drastically higher than ASB and MADN. 
 All three methods  established similar number of lightpaths, close to 1600 (the physical upper limit for 100 nodes carrying 16 transmitter/receivers).
 The quality of the solutions can be determined as the ratio of performance  to the number of lightpath changes, which we define as {\em efficiency}.
 Figure \ref{fixedEff} compares efficiencies of the three methods.
 
 Network operators are reluctant to make drastic  changes in their topologies, even if that means only changing the link weights \cite{Fortz:OOW02}.
 Thus, a VTR algorithm that can control the number of lightpath change is desirable. We aim for devising such an algorithm, and present the underlying probabilistic analysis.
 
 \begin{figure*}[th]
 %\begin{minipage}{1\textwidth}
 \centering
\def\mywidth{0.29}
    \subfloat[Under low traffic all methods perform similiar, while for heavy traffic HLDA performs best.]
    { \includegraphics[width=\mywidth\textwidth]{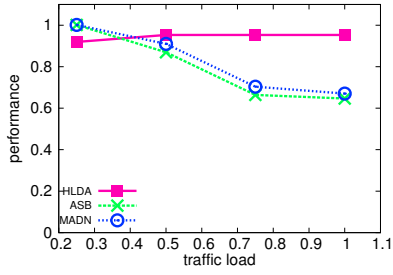} 
         \label{fixedPer}}
  \hfil
  \subfloat[HLDA changes lighthpaths much more than ASB and MADN per round. ] {\includegraphics[width=\mywidth\textwidth]{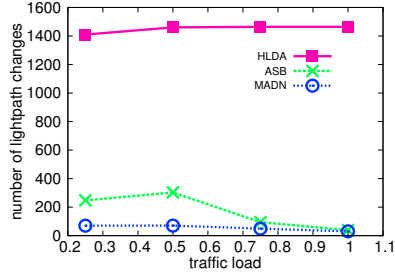} 
      \label{fixedLP}}
    \hfil \subfloat[MADN is the most efficient of all three. ]
    { \includegraphics[width=\mywidth\textwidth]{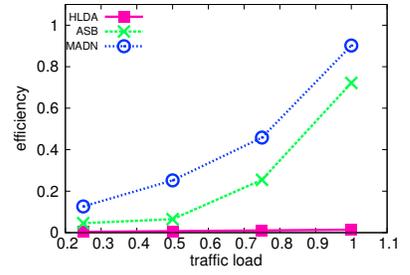} 
  \label{fixedEff}   } 
 % \end{minipage}
    \caption{ The performance comparison of various VTR algorithms.}
   \label{fixedVTR}
 \end{figure*} 
 
% \begin{itemize} 
%  \item   Heuristic Logic Topology Design Algorithm (HLDA)\,\cite{Ramaswami:DLTW96}.
%  \item Attractor Selection Based (ASB) method\,\cite{Koizumi:10}.
% \end{itemize}
\subsection{Comparison Method}
We chose HLDA, because it is an efficient heuristic and it is designed to minimize congestion, like ASB.
Note that, MILP based methods are unable to simulate large topologies (i.e. 100 nodes), thus we are bounded to use heuristic
methods.

Some other efficient methods aim to minimize single-hop traffic, end-to-end delay.
ASB is selected because its analysis is straightforward as we show in Section \ref{optimality}.
Although there are more efficient methods than ASB such as Multistate Attractor Selection with Dependent Noise (MADN) \cite{Hanay:VTCM13,Hanay:NTS14};
such methods are generally intractable. Our goal is not to evaluate the performance of different VTR algorithms. 
Instead, we are interested in exploring the controllability of VTR algorithms.
We explored the existence of a optimal control parameter for ASB.

\section{Preliminaries}\label{background}
ASB uses traffic loads of the links and HLDA uses the traffic matrix.
We assume that traffic loads are continuously monitored by a central controller for fixed intervals, 
and this is easy to implement in practice using Simple Network Management Protocol (SNMP) \cite{Wu:ASWN11}.

The following sections review the relevant part of ASB briefly, for a more thorough explanation of these two methods, reader can refer to
the corresponding paper \cite{Koizumi:10}, \cite{Ramaswami:DLTW96}. Yet, the following section should be self-sufficient to follow the discussion. 
After the overview of those two methods, 
we then explain how these two methods can be made tunable.  

\begin{figure}
   \setcounter{figure}{0}
  \renewcommand\figurename{Algorithm}
 \fbox{\begin{minipage}{0.95\columnwidth} 
\begin{algorithmic}[1]
\Procedure{ASB}{\,time t\,} 
%\State $r\gets a\bmod b$
%\While{True} 
%\State $V_G \gets  1/(e-^{GetMaxUtilization()})$  \Comment{Calculate $V_G$  }
 %\State $u_{max} \gets )$
%\State $V_G \gets CalculateVg(GetMaxUtilization())$  
\State $V_G \gets ComputeVg()$ \Comment{using $u_{max}$ by Eq. \ref{activityvsmax} }
\State $ComputeWeightMatrix(V_G, t)$
\State $ComputeExpression()$
\State $UpdateLightPath()$
%\EndWhile\label{euclidendwhile}
%\State \textbf{return} $b$ %\Comment{}
\EndProcedure

\\
\Procedure{ComputeWeightMatrix}{\,$V_G$, time t}
 
\If { $( V_G(t-1) < T_{max}\: \&\: V_G(t) > T_{max})$} 
  \For {$i \leftarrow 1, n$}
    \For  {$j \leftarrow 1, n$}
	\State $weightMatrix[i,j]\, \mathsmaller{-=}\, Hebb(i,j,A_k)$
	\EndFor
\EndFor
 \For {$i \leftarrow 1, n\times(n-1)$} 
  %\hskip 1.6cm // Update attractor matrix.    
    \State $ A_k[i] = LightPath[i]$ \Comment{Update attractors}
  \EndFor
 \For {$i \leftarrow 1, n$}
    \For  {$j \leftarrow 1, n$}
	\State $weightMatrix[i,j]  \, \mathsmaller{+=}\,Hebb(i,j,A_k)$
	%\State $AddToAttractorList()$
\EndFor
\EndFor

\State $k=(k+1)\mod numberOfAttractors $
\EndIf
  
\EndProcedure
\\
\Function{ComputeExpression}{}     
 \Comment{by Eq. \ref{furusawaEq}  }
  \For {$i \leftarrow 1, n \times (n-1)$}
    \For  {$j \leftarrow 1, n \times (n-1)$}
%	\State $x[i\times (n-1) + j]\, \mathsmaller{+=}\,CalcDeltaExp(i,j) $ 
             \State $x[i]\, \mathsmaller{+=}\,ComputeDeltaExp(i,j) $ 
\EndFor
\EndFor
\EndFunction
 \\
\Procedure{UpdateLightPath}{}
 %\If { $ V_G(i-1) < T_{max} \& V_G(i) > T_{max})$} 
  \For {$i \leftarrow 1, n\times(n-1)$}
	 \If { ($ x[i]>0.5 \: \&\: CanEstablish(i) ) $}
  	\State $EstablishLighpath(i)$ \Comment{LightPath[i]=1}
        \ElsIf { ($IsEstablished(i) \: \&\:  x[i]<0.5 ) $}
           \State $RemoveLighpath(i)$ \Comment{LightPath[i]=0}
	\EndIf
\EndFor  
\EndProcedure
 
\end{algorithmic}
% \end{algorithm}

\end{minipage}}

\caption{ASB method}
  \label{alg:algorithm}
\end{figure}

\subsection{Attractor Selection Based VT Control}\label{attbasedVNT}
ASB method searchs for topologies randomly, and if a satisfactory topology is found, it is 
saved in a memory. This saved topology is called an ``attractor''.
ASB method is described in Algorithm \ref{alg:algorithm}.
 %At first, the system either   generates attractors randomly or loads a list of attractors. 

 At the beginning of a round, the ASB controller 
 gets the utilization of the maximally loaded link and computes a performance  metric $V_G$, which is given below.
  \begin{equation}\label{activityvsmax}
 V_G = \frac{1}{1+e^{50(u_{max}-0.5)}}
\end{equation} 
In this equation, $u_{max}$ represents the utilization of the maximally loaded link.
 %, which is referred as ``activity'' in \cite{Koizumi:10}. 
 Before getting  into further details of the algorithm, we give some  numerical examples to clarify the meaning of the variables.
 Assume that there are 100 nodes, $n=100$, then the possible number of pairs is 9900, $n\times(n-1)$. Thus a topology can be described by a bit vector of size 9900.
For example, the first virtual topology in Figure \ref{vnt} can be described by the following bit vector:
\begin{align}
 [ \underbrace{0 1 0}_\text{node 1}\:\:  \underbrace{0 1 1}_\text{node 2}\:\:  \underbrace{ 1 1  1}_\text{node 3}\:\:  \underbrace{ 0 1 1}_\text{node 4} ] 
\end{align}
where a 1 bit indicates that the corresponding pair has a lightpath between them. For example, since node 3 is connected to all nodes, the bits belonging to node 3 are set to 1's.
 
%An attractor is a topology that satisfies some congestion requirements. 
The attractors are stored in an attractor matrix $A$.
For example, to store 5 topologies, the size of the attractor matrix $A$ must be 5 by 9900. 
$A_i$ denotes the $i^{th}$ attractor and the attractors are added in a FIFO sense (line 17).

After calculating $V_G$, the system compares whether the performance improved with respect to previous round significantly 
by checking against a threshold parameter $T_{max}$ as shown in line 8. 
If so, the topology is added as a new attractor by the means of changing the weight matrix as shown in line 16. $Hebb$ function
calculates the weights based on Hebb learning, which is given by Equation \ref{hebbeq}.
This new weight matrix generates new {\em expression levels}, $x$, which can be thought as a measure of how likely a lightpath has to be established.
For each lightpath $l_i$, there is a corresponding expression level $x_i$.

\begin{figure}[t] 
\centering
  \includegraphics[width=1\columnwidth	]{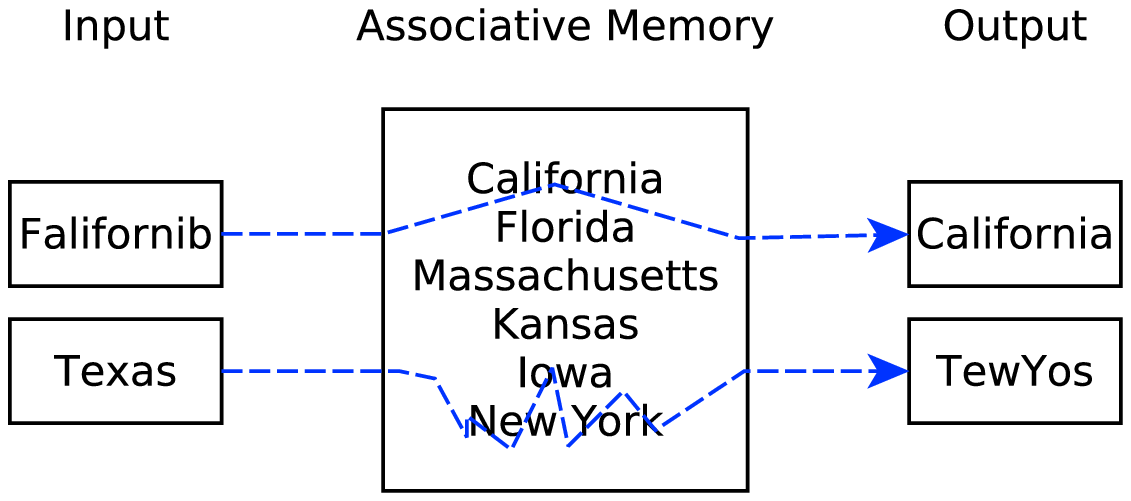}
 % activity.eps: 0x0 pixel, 300dpi, 0.00x0.00 cm, bb=201 50 554 554
\caption{ Auto-associative memories can recognize and correct noisy inputs to some extent.
A noisy input Falifornib is supplied, and the memory returns the closest element California.
On the other hand, if a non-existent entry (i.e. Texas) is presented, then a permutation of some stored input patterns is returned as the output. 
%Note that this is a conceptual example, and the memory holds the 6 states as a weight matrix.
}
\label{auto_asso}
\end{figure}

{\bf Auto-associative Memories}:
Auto-associative memories are neural memories, which are used to store patterns  based on a correlation matrix \cite{Kohonen:CMM72}.  
Neural memories are different than the computer memories; the values are not read in neural memories, they are calculated.
Auto-associative memories are conceptually similar to content-addressable memories (CAMs). 
To query the memory, user provides a data word instead of an address.
In addition, unlike computer memories, neural memories are not physical devices. 
The values are not read from bitcells,  but rather they are ``calculated'' by a matrix multiplication.
The concept of auto-associative is illustrated in Figure \ref{auto_asso}.
  The attractors are stored in an auto-associative memory.
  
In line 23, $ComputeDeltaExp$ function calculates the  $\frac{dx_i}{dt}$ according to Equation \ref{furusawaEq}.
The following equation shows how the expression level is updated \cite{Koizumi:10}: 
{  \begin{equation} \label{furusawaEq}
\frac{dx_i}{dt}= \underbrace{ \left[ f\left( \sum_{j=1}^n w_{ij} x_j   \right) - x_i \right]}_\text{auto-associative memory} V_G\: +\: 
 \underbrace{\vphantom{  f\left( \sum_{j=1}^n w_{ij} x_j   \right)  } \mathcal{N}(0,1)}_\text{random walk}
 \end{equation}  } 

 In this equation, $\mathcal{N}(0,1)$ is the standard normal random variable, and  $x_i$  captures the importance of a lightpath. 
If $x_i$ is greater than $0.5$,
 a lightpath is established provided that there are enough resources. A higher   $V_G$ means the system is in better condition. 
 %$\eta$ is zero mean unit-variance Gaussian noise. 
 $f(.)$ is the sigmoid function. 
  
The system dynamics shown in Eq. (\ref{furusawaEq}) consist of two components: auto-associative memory and random walk.
% The deterministic  component consists of the and the stochastic component is  random walk. 
When the system is in good conditions,
that is when $V_G$ is high, then the $x_i$ is mostly determined by the auto-associative memory,
which inclines towards to the stored memory elements.
On the contrary, when the network conditions get  worse and the controller needs to find a new attractor suitable for the new conditions, 
it randomly  searches for a new attractor. 

ASB does not make any assumption about network resources availability. 
If there are not enough resources for a lightpath to be established,
 ASB does not establish the lightpath, and skips to the next lightpath. 
An alternative approach is to continue looking for a topology  in which all the lightpaths can be established.
There are several ways to build  the weight matrix using different learning algorithms, such as Hebbian, Oja and APEX learning; 
and the effects of different learning algorithms has been studied before \cite{Hanay:VNT012}.
We use Hebbian learning for constructing weight matrix. 

{\bf Hebbian Learning}:
ASB's auto-associative memory uses Hebbian learning to store and read the elements. In our case, virtual topologies are stored 
in a auto-associative memory of which weight matrix is constructed using Hebbian learning.
 
In the general case, the weight matrix $W$ is constructed for a topology vector
X, as $W=X^{T} X$. Weight matrix can be constructed, using Hebbian learning weight update rule below:
\begin{equation}\label{hebbeq}
 \Delta w_{i,j} = {{\alpha}} l_{i} l_{j} 
\end{equation}
%and $W$ becomes the auto-correlation matrix.
Here, $\alpha$ is 
learning rate, which was set to 1. % $l_i$ is the lightpath indicator variable, and 
The weight matrix is updated when an attractor is found. 
To speed up the calculation, instead of a 9900 by 9900 matrix multiplication, we first subtract the contribution of the oldest attractor(line 11), and add the new one (line 16).
%Before the addition of the new attractor, the oldest attractor in the memory is removed 
\section{Controllability and Optimality}\label{methodology}
In the following sections we discuss tuning of the two algorithms, and present the analysis on optimizing search space for ASB.
 
\subsection{Controllability}
Most of the methods presented previous are not tunable.  
The most distinctive example is TILDA \cite{Ramaswami:DLTW96}, which assigns lightpaths based 
on the hop distance. TILDA is traffic-agnostic, that is, regardless of the traffic it generates the same virtual topology
as long as the physical topology stays same. TILDA can be made tunable  in several ways.
For example, an operator can modify TILDA so that the lightpaths assigned only between nodes that have a hop distance less than some specific number $h$.
It is possible to find a  sub-optimal $h$ by empirical analysis, but it is rather intractable to analyze mathematically.
Thus, although it is possible to make any given method {\em tunable}, 
the analysis of the control parameter is not straightforward.

The VTR method {\em Adaptive} presented in \cite{gencata:virtual-topology} can be considered as one of the first tunable method.
Adaptive uses two watermarks to establish a lightpath: $W_H$ and $W_L$. 
It is easy to minimize the search space for Adaptive, by setting $W_H =100$ and $W_L=0$, but it is not 
possible to analytically calculate how to maximize the search space (i.e. for which $W_H$ and $W_L$ values search space is maximized).
We can only say the that search space is maximized when $W_H=W_L$, but at which value this will be minimized depends on the traffic.

%A polynomial time algorithm was proposed under the assumption that lightpath are unidirectional and forms a ring \cite{errel:2006}.

\subsubsection{Tunable HLDA (tHLDA)}
We tried two different approaches to make HLDA tunable. In the first approach, the maximum 
number of lightpaths HLDA is fixed. In the other approach, the minimum amount of traffic that is able to establish a lightpath
is changed. In terms of efficiency and performance, the first method performed better, and we use this version of HLDA in this work. 
 
\subsubsection{Tunable ASB (tASB)}
In ASB, the noise term is a normal random variable with zero mean and unit variance. We make ASB by tunable by changing the noise term as follows:
\begin{equation}
\underbrace{   \mathcal{N}(0,1)}_\text{ASB}   \rightarrow \underbrace{ \mathcal{N}(\mu,1) }_\text{tunable ASB}
\end{equation}
Thus,  $\mu$ becomes a parameter for ASB. 
By looking Equation \ref{furusawaEq}, we can see that $\mu$ has an effect on $x_i$.
Statistically, a positive $\mu$   increases the value of $x_i$ and, a negative $\mu$ decreases the value of $x_i$ with respect to $\mu=0$ (original ASB). 
Since VTR is a NP-complete problem, an optimal $\mu$ minimizing the congestion cannot be found in polynomial time.
However, when a good topology is found for some $\mu$ value, it can be changed incrementally in either direction by removing or adding multiple lightpaths at once.
Note that changing $\mu$ has overall effect on all lightpath establishments and deletions.
This approach should not be confused with a pseudo-tuning strategy of other heuristic based methods, where the controller tries to increase the lightpath one by one. 
Here, in our method, $\mu$ can have different effects based on traffic and resources. In order to find an optimal $\mu$ value, we present our analytical calculation in the next section.
\subsection{Optimality}\label{optimality}
In this section, we present our analytical approach to calculate an optimum $\mu_{opt}$.
Some definitions that will be used in this section are presented in Table \ref{definitions}.
\begin{table}[h]
\caption{Definitions for variables and expressions.} \label{definitions}
\begin{tabular}{ll} \hline
%$p_{i-j}$ & The lightpath between node $i$ and $j$.\\
$P_i(j\rightarrow k, t)$ & Probability of lightpath $i$ changes from $j$ to $k$ at time t.	\\
$P_i(0\rightarrow 1, t)$ & Probability of  establishment of lightpath $i$ at time t.	\\
$P_i(1\rightarrow 0, t)$ & Probability of  termination of lightpath $i$  at time t.	\\
$X_{l_i}$	& resource availability Indicator random variable for $l_i$.				\\
\hline
 \end{tabular}
\end{table}
 
 \begin{figure*}[t]
  \centering
\def\factor{0.3}
    \subfloat[tHLDA performs better than tASB as the control parameter increases.] {\includegraphics[width=\factor\textwidth ]{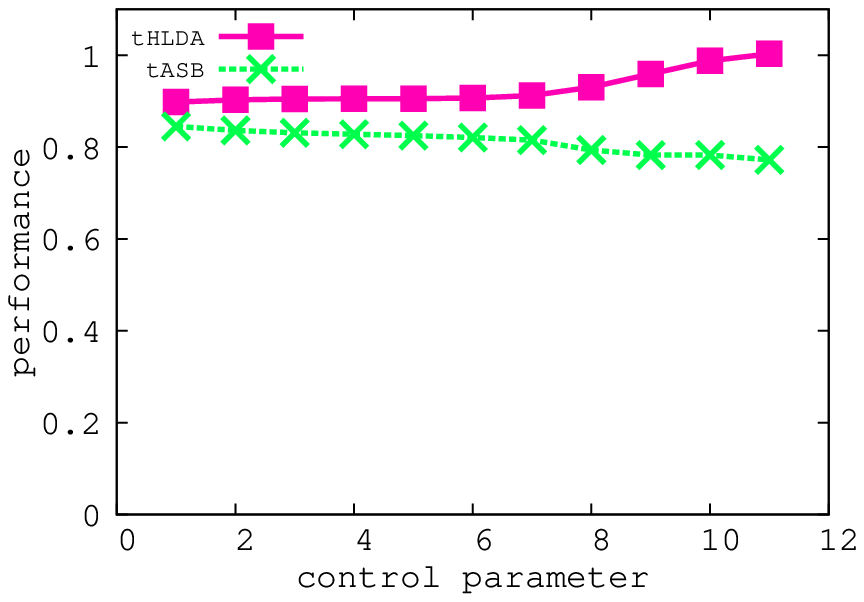} \label{diffloadlight}}   
\hfill
    \subfloat[The average number of lightpath changes per round increases for tHLDA, but it reduces for tASB. ]  { \includegraphics[width=\factor\textwidth ]{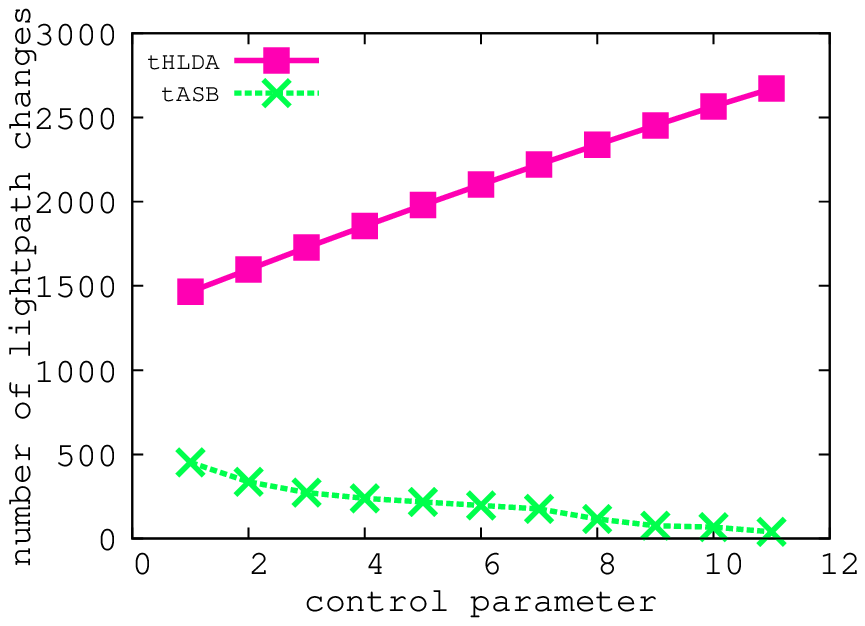} \label{reconf2}} 
\hfill
  \subfloat[tHLDA has almost a constant efficiency while the efficiency of tASB can be increased with the control parameter. ]{ \includegraphics[width=\factor\textwidth ]{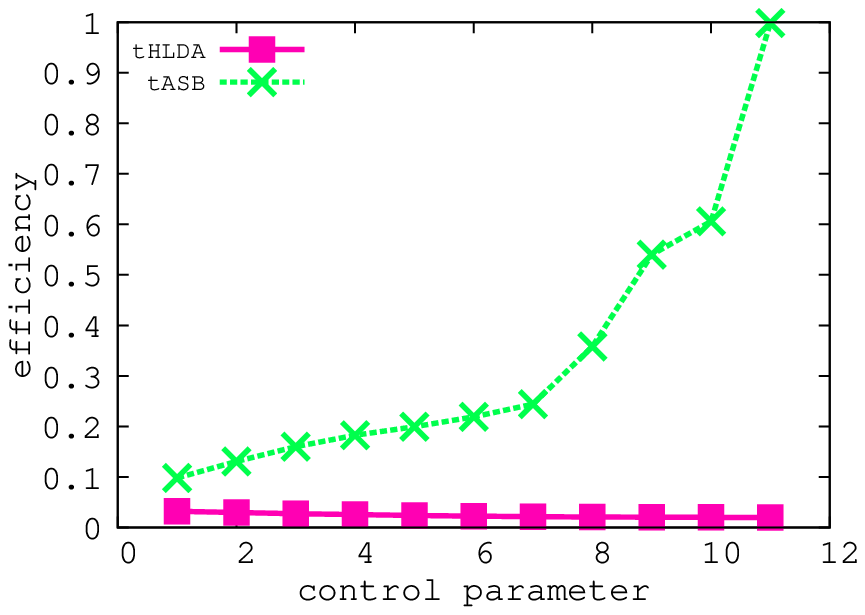} \label{reconf3}} 
\hfill
 
  \caption{ The comparison of tASB and tHLDA.} 
\label{reconf}
\end{figure*} 
 
One problem with ASB is that when the network conditions are poor, its new topology finding capability is limited.
Specifically, the network is performing poorly when $V_G = \epsilon$ for some small $\epsilon \approx 0$.
When the network conditions are poor, the probabilities of lightpath changes can be calculated by 
\par\nobreak \vspace{-0.24cm}
{\small \begin{align}
\label{probs2}
  &P(f(x_i):{0\rightarrow1}, t)& = \: &  1- P ( \eta < 0.5 \,|\, x_i(t-1)<0.5 ) \\
 &P(f(x_i):1\rightarrow0, t)& = \:&  P ( \eta < 0.5 \,|\, x_i(t-1)\geq 0.5 ) \\
&P_i(0\rightarrow 1, t) & =\: &   P(x_i:1\rightarrow0, t \,|\, X_{p_i} ) \\
&P_C & =\: &  P_i(0\rightarrow 1, t) \cup P_i(1\rightarrow 0, t) 
 \end{align}}%
Above, $P_C$ corresponds to probability of a lightpath addition or deletion.
Notice that in the equation we used the expression level, $x_i$, instead of the path indicator variable, $p_i$. We can analyze the probability of lightpath establishment and deletion using probability transitions.
The transition probability matrix $\mathcal{T}$ is defined as:
  \begin{align} 
\mathcal{T} =   \left[\begin{matrix} P(0\rightarrow 0) & P(0\rightarrow1)\\ P(1\rightarrow 0) & P(1\rightarrow1) \end{matrix} \right] 
\end{align}   
Note that we dropped the subscript $i$ and input $t$ from $P$ for abrevity, since the probability transitions are same for all paths and 
 rounds when $V_G=0$. For ASB,   we can calculate the transition probabilities for $V_G=0$, for which  $\mathcal{T}$ becomes: 
   \begin{align}\label{originalmat}
 \mathcal{T}_{ASB} =   \left[\begin{matrix} 0.69 & 0.31 \\ 0.69  &0.31 \end{matrix} \right] 
\end{align}  
 
When the $V_G$ is low, the VT controller has to find a new topology that satisfies the new traffic. 
If the traffic variance is high enough, then a topology that is much different than the last satisfactory topology must be found. 
In order to find such a diverse topology, the topology search space has to be increased. In the extreme case, the search space has to be maximized.
The search space can be maximized by making the transition probability matrix $\mathcal{T}=[0.5, 0.5; 0.5, 0.5]$, so that at any time the probability of lightpath establishment and deletion is equal.
% For $V_G=0$, this can be achieved by setting $\mu=0.5$ in the noise term in Equation \ref{furusawaEq}, so that
the noise term becomes $\mathcal{N}(0.5,1)$.
However, if $V_G > \epsilon$, the deterministic term in Equation \ref{furusawaEq} has to be considered also. 
We proceed by calculating probabilities for $V_G > 0$ by considering each transition separately. 
 \begin{align} \label{proeqs}
P(0\rightarrow0) &=  P(\eta < 0/5) \\  
P(0\rightarrow1) &=  P( Vg + \eta >0.5) \\
P(1\rightarrow0) &=  P (- Vg + \eta < 0.5) \\
P(1\rightarrow1) &=  P(\eta>0/5)   
  \end{align}  
In order to maximize the topology search space, probabilities in Equation   \ref{proeqs}  must be equal to $0.5$. 
In other words, we need to find the $\eta$ which would make all these probabilities equal to $0.5$ when $V_G=0$.
 Such a requirement is satisfied with setting   $\mu$ as below in each case:
  \begin{align}
  \centering
  \mu_{0\rightarrow0}  =\mu_{1\rightarrow1} & =  \hskip 0.46cm    0.5   \\
  \mu_{0\rightarrow1}& =    \left\{ \begin{array}{ll}
   0.5, &\mbox{ if $V_G=0$} \\
   0, &\mbox{ if $V_G>0.5$}
       \end{array} \right. \\
 \mu_{1\rightarrow0} & =  \left\{ \begin{array}{ll}
  0.5, &\mbox{ if $V_G=0$} \\
  1, &\mbox{ if $V_G>0.5$}
       \end{array} \right.\end{align}         
Combining all these equations together, we can write piece-wise linear functions between boundaries for each $\mu$.
The optimum mean, $\mu_{opt}(t)$ is found by 
 \begin{align} \label{uopt}
\mu_{opt}(t ) =\: &  \Big[ 0.5\: (1-x_i(t-1)) \: (1-x_i(t)) \Big] \: + \\ &  \Big[0.5 \: x_i(t-1)\: x_i(t) \Big] \:+ \nonumber\\
&  \Big[ (0.5 + V_G)  \: x_i(t-1) \: (1-x_i(t))  \: f(0.5-V_G)\Big] \: + \nonumber \\ &\Big[(0.5 - V_G) \:(1-x_i(t-1)) \: x_i(t)  \: f(0.5-V_G) \Big]\nonumber
\end{align}  
% \begin{equation} \label{proessqass2}
%    \begin{array}{l l l}
% \mu_{opt}(t ) =\: &  \Big[ 0.5\: (1-x_i(t-1)) \: (1-x_i(t)) \Big] + \\ &  \Big[0.5 \: x_i(t-1)\: x_i(t) \Big] +  \\
% &  \Big[ (0.5 + V_G)  \: x_i(t-1) \: (1-x_i(t))  \: f(0.5-V_G)\Big] + \nonumber \\ &\Big[(0.5 - V_G) \:(1-x_i(t-1)) \: x_i(t)  \: f(0.5-V_G) \Big] 
% \end{array} 
% \end{equation}
\normalsize
In Equation \ref{uopt}, the first term is for $(0\rightarrow0)$, the middle terms are for $(1\rightarrow1)$ and  $(1\rightarrow0)$, and the last term is for the $(0\rightarrow1)$ transition.

The discussion above assumes that network resources can over-provision the network. On the contrary, when the network resources cannot over-provision the traffic, the network resources need to be considered.
Instead of setting the  lightpath establishment probability to $0.5$, the availability of the network resources should be taken into account as below.
 \begin{equation}
P_i(0\rightarrow 1, t)=\frac{min(num_{ports},num_{wavelength})}{n-1} 
\end{equation} 
Here $num$ denotes the number of resources. $\mu_{opt}(t)$ can be calculated similarly for this new $P$. 
 
\section{Simulations} \label{simulations}
In this section we present our simulations results. 
The network consisted of 100 nodes, and the log-normal traffic model was used \cite{Nucci:PSGI05}.
For ASB method, an initial list of attractors was generated randomly. % and all the configurations used the same set of attractors.
Dijsktra's shortest path algorithm was used for routing, and the lightpaths were assumed to  be unidirectional. We assumed that the nodes are equipped with wavelength converters.
 The physical topology has 100 nodes, and its graph characteristics are given in Table \ref{physicaltab}.
 
 \begin{table}[h]
\caption{Physical topology characteristics}
\centering
\begin{tabular}{cccc} \hline   \label{physicaltab}
   Degree &          Avg. Path Length & Clus. Coeff. &Diameter \\ \hline 
 	4    	&	3.41	& 0.05 					&	6		 \\	
\hline
\end{tabular}
\end{table}

%100 nodes correspond to 9900 node pairs, thus $W$ is  9900 by $m_a$. 
 
First we compared tHLDA and tASB for different control parameters. 
The control parameter in tASB is $\mu$ which changes from $0$ to $0.5$, and the control parameter in tHLDA is 
the number of lightpaths which was set betwen 800 to 1600, with an increment of 80.  
The lower bound 800 was chosen as this was the point where tHLDA and tASB performance was equal.
For each control parameter, each method was run 30 times, and the average of those runs were taken.
Figure \ref{diffloadlight} and \ref{reconf2} show the performance and overhead comparisons. 
s the control number increases tHLDA outperforms tASB. Hovewer, Figure \ref{reconf2} 	
reveals that as the control parameter increases the number of lightpath changes is drastically higher 
than tASB again. More importantly, Figure \ref{reconf3} reveals that for HLDA, the tunability is quite low.
The figure shows that the number of lightpath change is also a fraction of total number of lightpaths.
The efficiency stays constant across all the control parameters.

%On average, lightpath change is $1.74$, and for r hb

\begin{figure}[hb] 
 \centering
   \includegraphics[width=1\columnwidth, ]{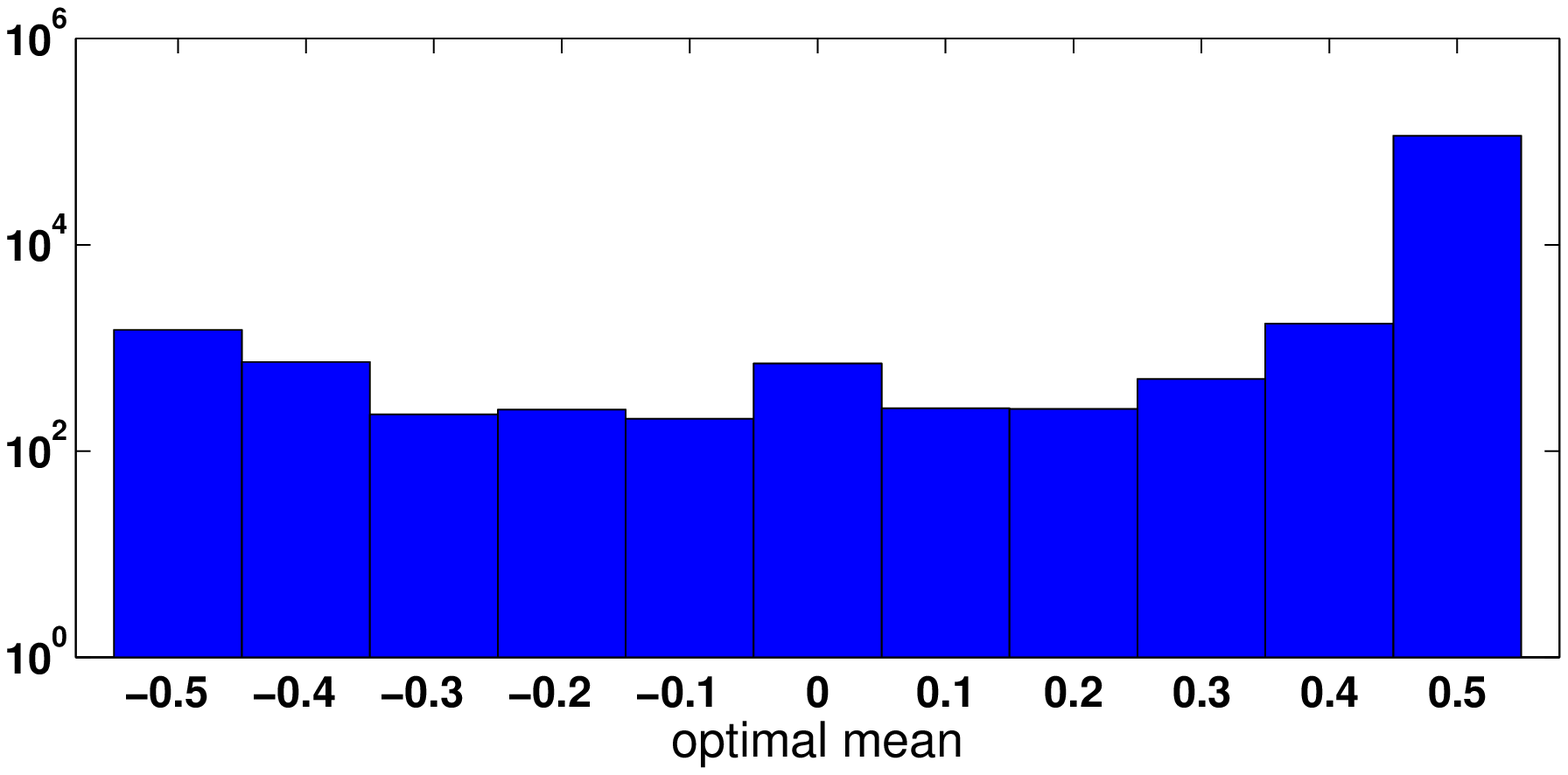}
 % activity.eps: 0x0 pixel, 300dpi, 0.00x0.00 cm, bb=201 50 554 554
\caption{The distribution of $u_{opt}$ values are concentrated around $0.5$.
 Note that the y-axis is log-scale.
 }
\label{opthist}
\end{figure}

In the second part, $\mu$ parameter was swept from $0.2$ to $0.6$.
Each configuration (for each $\mu$ value) was run with 10 random traffic patterns, and  the mean of 10 runs was taken. 
 $\mu_{opt}$ was sampled in each of these runs, for observation. 
We observed that $\mu_{opt}$ has a mean of $0.46$, with a standard deviation of $0.13$. 
The histogram of  $\mu_{opt}$ was given in Figure \ref{opthist}. 
It shows that $\mu=0.5$ appears 100 times more frequently than $\mu=0$.
%Thus, the effect of the memory term in Equation \ref{furusawaEq} can be 

\begin{figure*}[thb]
\centering
 \def\mywidth{0.3}
  % \begin{minipage}{1\textwidth}\centering
     \subfloat[Optimal mean value vs ASB performance. As the traffic load increases, VT controller with $\mu_{opt}$ outperforms ASB as expected.]
     {\includegraphics[width=\mywidth\textwidth]{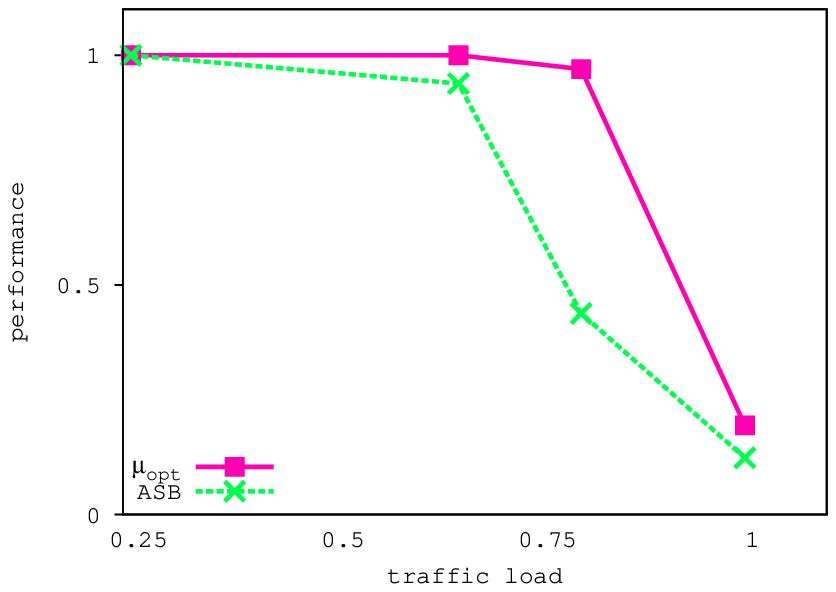}  \label{optcomp} }
 % activity.eps: 0x0 pixel, 300dpi, 0.00x0.00 cm, bb=201 50 554 554
 \hfill
        \subfloat[tASB performs best for $\mu=0.4$ for any traffic load.]     {  \includegraphics[width=\mywidth\textwidth]{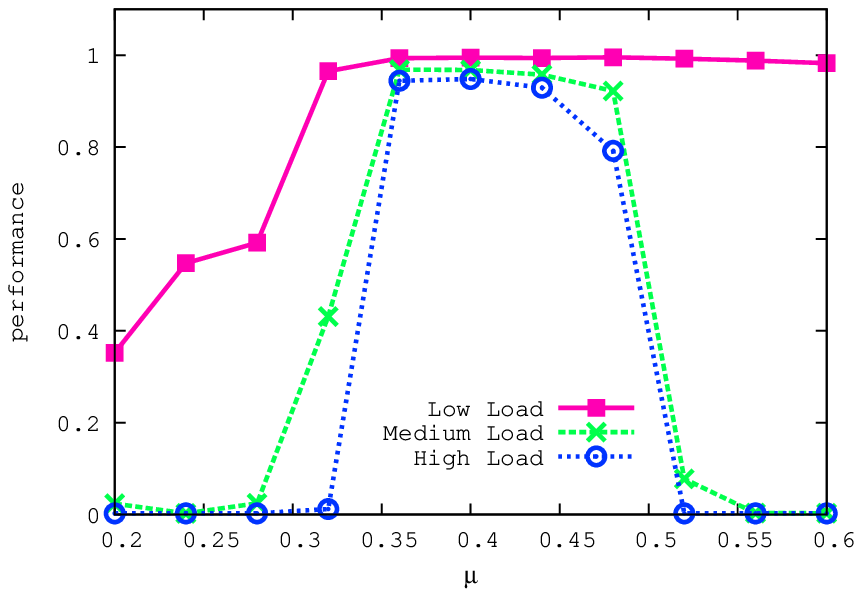} \label{constact}}
  \hfill
 % activity.eps: 0x0 pixel, 300dpi, 0.00x0.00 cm, bb=201 50 554 554
 \subfloat[The minimum number of lighpaths established when $\mu=0.32, \mu=0.36, \mu=0.38$ for low, medium and high traffic loads.]
{\includegraphics[width=\mywidth\textwidth ]{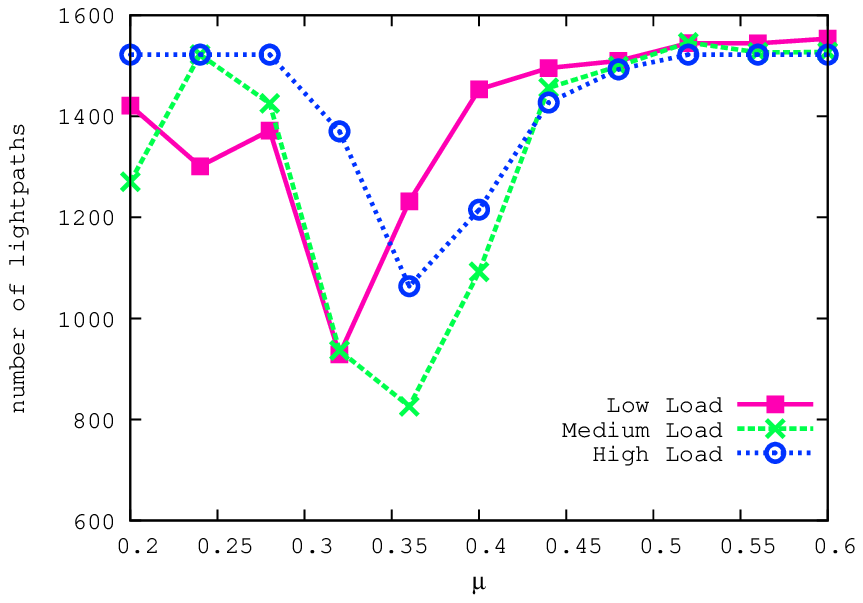} \label{constlight}}
 % activity.eps: 0x0 pixel, 300dpi, 0.00x0.00 cm, bb=201 50 554 554
 \label{diffload}
 \caption{Optimality of $\mu$  for various traffic loads.
 }
\end{figure*}

Figure \ref{optcomp} shows the comparison of ASB vs our extended analytical model with $\mu_{opt}$.
The figure indicates tASB with $\mu{opt}$ performs better than ASB.
Thus it is safe to say, our analytical findings about optimal $\mu$ agrees with our simulations.
The simulations were run for 400 steps. We conjecture that increasing simulation time would  increase the performance further.
The main drawback of the $\mu_{opt}$ approach is its longer running time, which is about 10X slower than  tASB.
Thus we made another set of simulations to analyze how tASB performs under constant $\mu$ values.
  Figure \ref{constact} shows the performance of the controller with various $\mu$ values, under 
three different traffic patterns. As our analytical calculations suggested, $\mu < 0.5$ gives the best results. 

When the traffic load is low, higher $\mu$ values increases the value, as more lightpaths start to establish. 
However, for medium and high loads, increasing $\mu$ beyond $0.5$ results in poor performing topologies. 
 This is due to the sigmoid function, where $\mu=0.5$ is a saddle point. $\mu=0.5$ means that most nodes pairs start to have a lightpath assigned. 
Since lightpath assignment is random, less important paths deplete the resources and results in resource scarcity for more important lightpaths that are assigned later in lightpaths assignment.
This situation, results in use of all available lightpaths to be used as it can be seen in Figure \ref{constlight}.
Thus, every pair  experiences the same amount of increase, and a $\mu=0.5$ or higher is not meaningful.
%We also observed that traffic variance does not affect the performance.

Figure \ref{constlight} emphasizes the linear relation between $\mu$ and the number of lightpaths establishment. 
As the $\mu$ increases, $P(0\rightarrow1)$ increases, and the more lightpaths establish. This explains the behavior for $\mu>0.35$.
However, for $\mu<0.35$ configurations, the network initially starts with lower number of lightpaths.
Since its search space is small, it cannot find a good topology; and the controller incrementally increases the number of lightpaths. Throughout the simulation,
those configurations failed to find a good topology even when the number of lightpaths reached the maximum. 
Because the network fails to find a good topology for $\mu<0.35$ as can be seen in Figure  \ref{constact}.
For example, setting $\mu=0.4$ costs 40\% fewer lightpath establishments.
  The figure also shows  that the number of assignments reach to physical limit of 1600.

\section{Conclusion}\label{conclusion}
In this paper, we  presented tunable versions of HLDA and ASB 
and analyzed their tunability.
We showed that the tunability of tHLDA is   low, while tASB is markedly high.

Then, we analyzed for an optimum $\mu$ parameter for tASB, and we did simulations to 
observe how  it performs under various traffic.
Our motivation was based on the fact that most of the previous VTR methods do not have any parameters. 
In the previous methods that have parameters, the analysis of the parameters are typically limited   to empirical analysis.
This paper is the first to show an analysis on how to determine a range for parameter.
 
We conclude that for each traffic pattern and network configuration there is an optimal range of $\mu_{opt}$ value that would result in minimum lightpath changes and maximum performance.
Since our system dynamics capture all the lightpaths through $x_i$, it is more efficient and scalable than   heuristic methods where
tuning is achieved by sorting lightpaths based on their load level, and assigning light paths one by one.

We showed that main problem
with ASB is that it has a rather low probability to establish lightpaths (i.e. 0.31) when the network is in poor conditions. By solving for optimal $\mu$ value, 
our controller outperformed ASB with various traffic loads.
 
For future work we will work on some other VTR algorithms and we will consider the effect of  physical layer impairments
on the tunability.
\bibliographystyle{IEEEtran}
\bibliography{arxiv2015}

% that's all folks
\end{document}